\documentclass[a4paper]{article}
\usepackage{fullpage}
\usepackage{amsmath}

\begin{document}
\title{Analytic matrix elements with shifted correlated Gaussians}
\author{D.V.~Fedorov \\
\it Institute of Physics and Astronomy, Aarhus University, \\
\it Ny Munkegade 120, 8000 Aarhus C, Denmark \\
\it fedorov@phys.au.dk
}
\date{}
\maketitle

	\begin{abstract}
Matrix elements
between shifted correlated Gaussians
of various potentials with several form-factors
are calculated analytically.
Analytic matrix elements are of importance for the correlated Gaussian
method in quantum few-body physics.
	\end{abstract}

\section{Introduction}

Correlated Gaussian method is a popular variational method to
solve quantum-mechanical few-body problems in molecular, atomic,
and nuclear physics~\cite{mitroy,ChemRev}.  One of the important
advantages of the correlated Gaussian method is the ease of computing
the matrix elements. In some cases the matrix elements and even
their gradients with respect to optimization parameters are fully
analytic~\cite{ChemRev,cafiero-analytic,sharkey-analytic,DailyGreene}.
This enables extensive numerical optimizations to be carried out leading
to accurate results~\cite{tung-calc,bubin-calc,blume-semi} despite the
incorrect functional form of the Gaussians in certain asymptotic regions
of the configuration space.

Although a number of analytic matrix elements have been calculated for
different potentials and different forms of correlated Gaussians, one
combination --- short-range potentials with shifted correlated Gaussians
--- is still missing~\cite{mitroy}.  Indeed historically the shifted
Gaussians have been applied more often to Coulombic systems rather than
to atomic and nuclear systems where the short-range interactions are
most important.

In this paper several types of short-range potentials are considered in
a search for the form-factors that produce analytic matrix elements with
shifted correlated Gaussians.  A brief introduction to the correlated
Gaussian method is given first and then the analytic matrix elements
are introduced.

\section{Correlated Gaussian method}

Correlated Gaussian method is a variational method where the coordinate
part of the wave-function of a quantum few-body system is expanded in
terms of correlated Gaussians. Various forms of correlated Gaussians
have been concieved~\cite{mitroy}, one of them being the shifted
correlated Gaussian, $|g\rangle$, which for a system of $N$ particles
with coordinates $\vec r_i|_{i=1\dots N}$ has the form

	\begin{equation}
\langle \mathbf r|g\rangle = \exp\left( -\sum_{i,j=1}^{N} A_{ij}\,\vec
r_i\cdot\vec r_j + \sum_{i=1}^N\vec s_i\cdot\vec r_i \right) \equiv
e^{-\mathbf r^\mathsf TA\mathbf r+\mathbf s^\mathsf T\mathbf r} \,,
	\end{equation}
where $\mathbf r$ is size-$N$ column of particle coordinates $\vec r_i$,
$A=\{A_{ij}\}$ is a size-$N$ square symmetric positive-definite
{\em correlation matrix}, $\mathbf s$ is size-$N$
column of {\em shift vectors} $\vec s_i$, and where the
following notation has been introduced,

	\begin{equation}
\mathbf r^\mathsf TA\mathbf r \equiv \sum_{i,j=1}^{N} A_{ij}\,\vec
r_i\cdot\vec r_j \,,\quad
\mathbf s^\mathsf T\mathbf r \equiv \sum_{i=1}^N\vec s_i\cdot\vec r_i \,,
	\end{equation}
where ``$\cdot$'' denotes the dot-product of two vectors.  The elements
of the correlation matrix and the shift vectors are the non-linear
parameters of the Gaussians.

The coordinate part of the few-body wave-function
$|\psi\rangle$ is represented as a linear combination of several
Gaussians,
	\begin{equation}
|\psi\rangle = \sum_{i=1}^{n_g} c_i|g_i\rangle \,,
	\end{equation}
where the coefficients $c_i$ are the linear parameters, and where
$n_g$ is the number of Gaussians.
Inserting this representation into the Schrödinger equation,
	\begin{equation}
\hat H |\psi\rangle = E |\psi\rangle \,,
	\end{equation}
where $\hat H$ is the Hamiltonian of the few-body system, and multiplying from the left with
$\langle g_j|$ leads to the generalized matrix eigenvalue equation,
	\begin{equation}\label{eq:gev}
\mathcal H c = E\mathcal N c \,,
	\end{equation}
where $c=\{c_i\}$ is the column of the linear parameters, and where the
Hamilton matrix $\mathcal H = \{\mathcal H_{ij}\}$ and the overlap matrix
$\mathcal N = \{\mathcal N_{ij}\}$ are given as
	\begin{equation}
\mathcal H_{ij} = \langle g_i | \hat H | g_j \rangle \,,\quad \mathcal N_{ij}
= \langle g_i | g_j \rangle \,.
	\end{equation}

The linear parameters together with the energy spectrum are found by
solving the generalized eigenvalue problem~(\ref{eq:gev}) numerically 
using the standard linear algebra methods~\cite{gsl}.  The non-linear
parameters are optimized by one of many optimization methods which typically
involve elements of stochastic-evolutionary~\cite{suzukivarga} and
direct optimization algorithms~\cite{bubin-calc,blume-semi}.  The direct
optimization algorithms often employ gradients of the matrix elements
with respect to the optimization parameters.

These optimization techniques involve numerous evaluations of the
Hamiltonian matrix elements and their gradients.
Therefore the analytica matrix elements are of particular importance
for the method.

\section{Matrix elements}
\subsection{Overlap}
The overlap $\langle g'|g\rangle$ between a shifted Gaussian $|g\rangle$
with parameters $A$, $\mathbf s$ and a shifted 
Gaussian $|g'\rangle$ with parameters $A'$,
$\mathbf s'$ is given as\footnote{
The overlap can be evaluated by an orthogonal coordinate transformation,
$\mathbf r=Q\mathbf x$, to the basis where the matrix $B$ is diagonal:
$B=QDQ^\mathsf T$ where $Q^\mathsf TQ=QQ^\mathsf T=1$ and $D$ is a
diagonal matrix,
	\begin{eqnarray}\label{eq:a:o}
\langle g'|g\rangle
&=& \int d^3\vec r_1\dots d^3\vec r_N \exp\left(-{\mathbf r}^{\mathsf T}B{\mathbf r}+\mathbf v^\mathsf T\mathbf r\right)
= \int d^3\vec x_1\dots d^3\vec x_N
\exp\left(-\sum_{i=1}^N{\vec x}_i\cdot D_{ii}{\vec x}_i+\sum_{i=1}^N \vec v_i\cdot\vec x_i\right)
\\
&=& \prod_{i=1}^N \int d^3\vec x_i \exp\left(-{\vec x}_i\cdot D_{ii}{\vec x}_i+\vec v_i\cdot\vec x_i\right)
= \prod_{i=1}^N \exp\left(\frac{1}{4D_{ii}}\vec v_i^2\right)\left(\frac{\pi}{D_{ii}}\right)^{3/2}
= e^{\frac14\mathbf v^\mathsf T B^{-1}\mathbf v}\left(\frac{\pi^N}{\det(B)}\right)^{3/2} \;.
\nonumber
	\end{eqnarray}
}
	\begin{equation}
\langle g'|g\rangle
=
e^{\frac14\mathbf v^\mathsf TB^{-1}\mathbf v}\left(\frac{\pi^N}{\det(B)}\right)^{3/2}
\equiv M
\,,\;
	\end{equation}
where $B = A'+A$, $\mathbf v = \mathbf s'+\mathbf s$.

\subsection{Kinetic energy}
The non-relativistic kinetic energy operator $\hat K$ for an $N$-body
system of particles with coordinates $\vec r_i$ and masses $m_i$ is
given as
	\begin{equation}
\hat K=-\sum_{i=1}^N \frac{\hbar^2}{2m_i} \frac{\partial^2}{\partial\vec
r_i^{\,2}} \,.
	\end{equation}
For completeness we shall consider a more general form of the kinetic energy operator,
	\begin{equation}
\hat K=-\sum_{i,j=1}^N\frac{\partial}{\partial\vec r_i}\Lambda_{ij}\frac{\partial}{\partial\vec r_j}
\equiv
-\frac{\partial}{\partial\bf r}\Lambda\frac{\partial}{\partial\mathbf r^\mathsf T}
%-\mathbf\nabla_\mathbf r^\mathsf T \Lambda \mathbf\nabla_\mathbf r
\,,
	\end{equation}
where $\Lambda$ is a symmetric positive-definite matrix.
The matrix element of this operator is given as\footnote{
We first calculate two integrals,
	\begin{equation}
\left\langle g'\left| \mathbf r \right|g\right\rangle
= \left( \frac{\partial}{\partial\mathbf v^\mathsf T}\right)
e^{\frac14\mathbf v^\mathsf T B^{-1}\mathbf v}\left(\frac{\pi^N}{\det(B)}\right)^{3/2}
=\mathbf u e^{\frac14\mathbf v^\mathsf T B^{-1}\mathbf v}\left(\frac{\pi^N}{\det(B)}\right)^{3/2}
\,,
	\end{equation}
where $\mathbf u=\frac12 B^{-1}\mathbf v$, and
	\begin{eqnarray}\label{eq:rFr}
\left\langle g'\left| \mathbf r^\mathsf TF \mathbf r \right|g\right\rangle
= \left(\frac{\partial}{\partial\mathbf v}F\frac{\partial}{\partial\mathbf v^\mathsf T}\right)
e^{\frac14\mathbf v^\mathsf T B^{-1}\mathbf v}\left(\frac{\pi^N}{\det(B)}\right)^{3/2}
=\left(\frac32\,\mathrm{trace}(FB^{-1})+\mathbf u^\mathsf T F \mathbf u\right)
e^{\frac14\mathbf v^\mathsf T B^{-1}\mathbf v}\left(\frac{\pi^N}{\det(B)}\right)^{3/2}
\,,
	\end{eqnarray}
from which the sought integral,
	\begin{equation}
\left\langle g'\left|
-\frac{\partial}{\partial\bf r} \Lambda \frac{\partial}{\partial\mathbf r^{\mathsf T}}
\right|g\right\rangle
=
\left\langle g'\left|
(\mathbf s'-2A'\mathbf r)^\mathsf T
\Lambda
(\mathbf s-2A\mathbf r)
\right|g\right\rangle \,,
	\end{equation}
follows directly.
}

	\begin{eqnarray}
\left\langle g'\left|
-\frac{\partial}{\partial\bf r}
\Lambda
\frac{\partial}{\partial\mathbf r^{\mathsf T}}
\right|g\right\rangle
=
\left\langle g'\left|
(\mathbf s^{\prime\mathsf T}-2\mathbf r^\mathsf TA')
\Lambda
(\mathbf s-2A\mathbf r)
\right|g\right\rangle
\nonumber\\
=
\left(
6\,\mathrm{trace}(A'\Lambda A B^{-1})
+(\mathbf s'-2A'\mathbf u)^\mathsf T
\Lambda
(\mathbf s-2A\mathbf u)
\right)
M
\,,
	\end{eqnarray}
where $\mathbf u\doteq\frac12 B^{-1}\mathbf v$, and $M$ is the overlap.

\subsection{Potential energy}
\subsubsection{Central potential}

A one-body central potential, $V(\vec r_i)$, and a two-body
central potential, $V(\vec r_i-\vec r_j)$, can be written in a convenient general
form, $V(w^\mathsf T\mathbf r)$, where $w$ is a size-$N$ column of
numbers with all components equal zero except for $w_i=1$ for the one-body
potential and $w_i=-w_j=1$ for the two body potential.

\paragraph{Gaussian form-factor}
For the Gaussian form-factor,
$V(w^\mathsf T\mathbf r)\propto e^{-\gamma\mathbf r^\mathsf T ww^\mathsf T\mathbf r}$,
the matrix element directly follows from the overlap integral,

	\begin{equation}\label{eq:mprime}
\left\langle g'\left|
e^{-\gamma\mathbf r^\mathsf T ww^\mathsf T\mathbf r}
\right|g\right\rangle
=
e^{\frac14\mathbf v^\mathsf TB'^{-1}\mathbf v}
\left(\frac{\pi^N}{\det(B')}\right)^{3/2}
\equiv M'
\,,
	\end{equation}
where the matrix $B'$ is a rank-1 update of the matrix $B=A'+A$,
$B'=B+\gamma ww^\mathsf T$.

If the determinant and the inverse of the matrix $B$ are known, their rank-1 updates
can be calculated efficiently using the update formulas,
	\begin{equation}
\det(B+ab^\mathsf{T}) = (1 + b^\mathsf{T}B^{-1}a)\det(B)\,,
	\end{equation}
	\begin{equation}
(B+ab^\mathsf T)^{-1} = B^{-1} - {B^{-1}ab^T B^{-1} \over 1 + b^T B^{-1}a} \,,
	\end{equation}
where $a$ and $b$ are size-$N$ columns of numbers.

\paragraph{Oscillator form-factor}
Another potential with a simple analytic matrix element is the oscillator
potential, $V(w^\mathsf T\mathbf r)\propto \mathbf r^\mathsf T ww^\mathsf T\mathbf
r$, relevant for cold atoms in traps. The matrix element directly follows from~(\ref{eq:rFr}),
	\begin{equation}
\left\langle g'\left|
\mathbf r^\mathsf T ww^\mathsf T\mathbf r
\right|g\right\rangle
=
\left(
\frac32 w^\mathsf TB^{-1}w + \mathbf u^\mathsf T ww^\mathsf T\mathbf u
\right)
M
\,.
	\end{equation}

\paragraph{Other analytic form-factors}
For a potential with a general form-factor, $V\propto f(w^\mathsf T\mathbf r)$, the matrix element
reduces to a three-dimensional integral,\footnote{
Suppose the form-factor $f(\vec r)$ has a Fourier-transform $\mathcal F(\vec k)$, then
       \begin{equation}
\left\langle g'\left|
f(w^\mathsf T\mathbf r)
\right|g\right\rangle
=
\int\frac{d^3\vec k}{(2\pi)^3}\mathcal F(\vec k)
\left\langle g'\left|
e^{i\vec kw^\mathsf T\mathbf r}
\right|g\right\rangle
=
e^{\frac14\mathbf v^\mathsf TB^{-1}\mathbf v}
\left(\frac{\pi^N}{\det(B)}\right)^{3/2}
\int\frac{d^3\vec k}{(2\pi)^3}\mathcal F(\vec k)e^{-\alpha k^2 + i\vec k\vec q}
        \end{equation}
where $\alpha=\frac14w^\mathsf TB^{-1}w$, $\vec q=\frac12 w^\mathsf TB^{-1}\mathbf v$. Now the last integral can as well be
written as
	\begin{equation}
\left(\frac\beta\pi\right)^\frac32
\int d^3\vec r \, f(\vec r) \, e^{-\beta (\vec r-\vec q)^2} \,,
	\end{equation}
where $\beta=\frac1{4\alpha}=(w^\mathsf TB^{-1}w)^{-1}$.
}
	\begin{equation}
\left\langle g'\left|
f(w^\mathsf T\mathbf r)
\right|g\right\rangle
=
%e^{\frac14\mathbf v^\mathsf TB^{-1}\mathbf v}\left(\frac{\pi^N}{\det(B)}\right)^{3/2}
M
\left(\frac\beta\pi\right)^\frac32
\int d^3\vec r \, f(\vec r) \, e^{-\beta (\vec r-\vec q)^2} \,,
	\end{equation}
where $\beta = \left( w^\mathsf T B^{-1} w \right)^{-1}$ and $\vec q = w^\mathsf T \mathbf u$.

If the potential does not depend on the direction of its argument
the integral reduces further to a one-dimensional integral,
	\begin{eqnarray}\label{eq:mati}
\left\langle g'\left|
f(|w^\mathsf T\mathbf r|)
\right|g\right\rangle
&=&
%e^{\frac14\mathbf v^\mathsf TB^{-1}\mathbf v}\left(\frac{\pi^N}{\det(B)}\right)^\frac32
M
\left(\frac\beta\pi\right)^\frac32
2\pi\frac{e^{-\beta q^2}}{\beta q}
\int_{0}^{\infty} r dr \, f(r) \, e^{-\beta r^2}\sinh(2 \beta q r)
\\
&\equiv& M J[f]
\nonumber\,,
	\end{eqnarray}
where
	\begin{equation}\label{eq:J}
J[f]\doteq
\left(\frac\beta\pi\right)^\frac32
2\pi\frac{e^{-\beta q^2}}{\beta q}
\int_{0}^{\infty} r dr \, f(r) \, e^{-\beta r^2}\sinh(2 \beta q r)
\,.
	\end{equation}

The integral~(\ref{eq:J}) gives relatively simple analytic
results for the Coulomb form-factor, $1/r$,

        \begin{equation}
J\left[\frac1r\right]
=
\frac{\mathrm{erf}(\sqrt\beta q)}{q}
\xrightarrow[q\to 0]{} \frac{2}{\sqrt{\pi}}\sqrt{\beta}
\,,
        \end{equation}
and for several short-range form-factors:
\begin{itemize}
\item screened Yukawa, $e^{-\gamma r^2-\mu r}/r$,
	\begin{eqnarray}\label{eq:syuk}
&&
J\left[
\frac{e^{-\gamma r^2-\mu r}}{r}
\right]
=
\\
&&
\frac{\sqrt{\beta } e^{-\beta q^2}}{2 q \sqrt{\beta +\gamma }}
\left(e^{\frac{(\mu -2 \beta  q)^2}{4
   (\beta +\gamma )}} \left(\mathrm{erf}\left(\frac{2 \beta  q-\mu }{2 \sqrt{\beta +\gamma
   }}\right)+1\right)+e^{\frac{(\mu +2 \beta  q)^2}{4 (\beta +\gamma )}}
   \left(\mathrm{erf}\left(\frac{\mu +2 \beta  q}{2 \sqrt{\beta +\gamma
   }}\right)-1\right)\right)
\nonumber \,,
	\end{eqnarray}

\item Yukawa, $e^{-\mu r}/r$,
        \begin{eqnarray}\label{eq:short1}
&&
J\left[
\frac{e^{-\mu r}}{r}
\right]
=
\lim_{\gamma\to 0}J\left[
\frac{e^{-\gamma r^2-\mu r}}{r}
\right]
=
\\
&&
\frac{e^{\frac{\mu^2}{4\beta}-\mu q}}{2 q}
\left(
1- e^{2\mu q} + \mathrm{erf}\left(\frac{2\beta q - \mu}{2\sqrt\beta}\right)
+e^{2\mu q}\mathrm{erf}\left(\frac{2\beta q + \mu}{2\sqrt\beta}\right)
\right)
\,,
        \end{eqnarray}

\item screened Coulomb, $e^{-\gamma r^2}/r$,
        \begin{eqnarray}\label{eq:short2}
J\left[
\frac{e^{-\gamma r^2}}{r}
\right]
=
\lim_{\mu\to 0}J\left[
\frac{e^{-\gamma r^2-\mu r}}{r}
\right]
=
\frac{\sqrt{\beta } e^{-\frac{\beta  \gamma  q^2}{\beta +\gamma }}
   \mathrm{erf}\left(\frac{\beta  q}{\sqrt{\beta +\gamma }}\right)}{q \sqrt{\beta +\gamma }}
\,,
        \end{eqnarray}

\item screened exponential, $e^{-\gamma r^2-\mu r}$,
        \begin{eqnarray}\label{eq:sexp}
&&
J\left[
e^{-\gamma r^2 - \mu r}
\right] 
=
\frac{-\partial}{\partial\mu}
J\left[
\frac{e^{-\gamma r^2 - \mu r}}{r}
\right]
=
\frac{-\sqrt{\beta }}{4 q (\beta +\gamma )^{3/2}}
e^{\frac{(\mu -2 \beta  q)^2}{4 (\beta +\gamma )}-\beta  q^2}
\\
&&
   \left((\mu -2 \beta  q) \mathrm{erf}\left(\frac{2 \beta  q-\mu }{2 \sqrt{\beta +\gamma
   }}\right)
+(\mu +2 \beta  q) e^{\frac{2 \beta  \mu  q}{\beta +\gamma }}
   \mathrm{erf}\left(\frac{\mu +2 \beta  q}{2 \sqrt{\beta +\gamma }}\right)
+\mu -2 \beta  q
   e^{\frac{2 \beta  \mu  q}{\beta +\gamma }}-\mu  e^{\frac{2 \beta  \mu  q}{\beta +\gamma
   }}-2 \beta  q\right)
\nonumber\,,
        \end{eqnarray}

\item exponential, $e^{-\mu r}$,
        \begin{eqnarray}\label{eq:exp}
&&
J\left[
e^{-\mu r}
\right]
=
\lim_{\gamma \to 0}
J\left[
e^{-\gamma r^2 - \mu r}
\right]
=
\frac{-e^{\frac{\mu  (\mu -4 \beta  q)}{4 \beta }}}{4 \beta  q}
\\
&&
\left((\mu -2 \beta  q)
   \mathrm{erf}\left(\frac{2 \beta  q-\mu }{2 \sqrt{\beta }}\right)+e^{2 \mu  q} (\mu +2 \beta
    q) \mathrm{erf}\left(\frac{\mu +2 \beta  q}{2 \sqrt{\beta }}\right)+\mu -2 \beta  q e^{2
   \mu  q}-2 \beta  q-\mu  e^{2 \mu  q}\right)
\nonumber
\,.
        \end{eqnarray}
\item Gaussian, $e^{-\gamma r^2}$,
	\begin{equation}
J\left[
e^{-\gamma r^2}
\right]
=
\lim_{\mu \to 0}
J\left[
e^{-\gamma r^2 - \mu r}
\right]
=
\left(\frac{\beta}{\beta+\gamma}\right)^{3/2}
e^{-\frac{\beta\gamma q^2}{\beta+\gamma}}
\,.
	\end{equation}
\end{itemize}

Since the error function has analytic derivative,
$\mathrm{erf}'(x)=2e^{-x^2}/\sqrt\pi$,
the following form-factors also have analytic matrix elements,
	\begin{equation}\label{eq:sydesc}
J\left[
r^n \frac{e^{-\gamma r^2 - \mu r}}{r}
\right]
=
\left(\frac{-\partial}{\partial\mu}\right)^n
J\left[
\frac{e^{-\gamma r^2 - \mu r}}{r}
\right]
\,.
	\end{equation}

Although there are few other form-factors with analytic matrix elements,
the resulting relatively complicated expressions are not conducive
to subsequent analytic calculations of gradients with respect to the
elements of matrices $A$ and the shift-vectors $\mathbf s$ through the
quantities $\beta$ and $q$.  In such cases it is probably more efficient
to represent the potential as a linear combinations of several Gaussians.

\subsubsection{Tensor potential}\label{ch:tensor}

The tensor potential $V_\mathrm t$ between two particles with coordinates
$\vec r_1$, $\vec r_2$ and spins $\vec S_1$, $\vec S_2$ can be written
in the form~\cite{ring}
	\begin{equation}
V_\mathrm t(r) \propto f(r)(\vec S_1\cdot\vec r)(\vec S_2\cdot\vec r) \,,
	\end{equation}
where $\vec r=\vec r_1-\vec r_2$ is the relative coordinate between the
particles, and $f(r)$ is the radial form-factor of the potential. In this
form the potential has a central spin-spin component,

	\begin{equation}\label{eq:cc}
\frac13f(r)r^2(\vec S_1\cdot\vec S_2) \,,
	\end{equation}
which is often subtracted from the above form to make sure the potential
contains only the spherical tensor component.

Introducing the size-$N$ column of numbers $w$,
	\begin{equation}
w=\{w_i|w_1=1, w_2=-1, w_{i\ne 1,2}=0\} \,,
	\end{equation}
and vector-columns $\mathbf y_1=\vec S_1w$, $\mathbf y_2=\vec S_2w$,
the tensor potential can be written in a convenient general form,
	\begin{equation}\label{eq:vtf}
\hat V_\mathrm t \propto f(w^\mathsf T\mathbf r)
(\mathbf y_1^\mathsf T\mathbf r)(\mathbf y_2^\mathsf T\mathbf r) \,,
	\end{equation}
where
	\begin{equation}
\mathbf y^\mathsf T\mathbf r
\equiv
\sum_{i=1}^N \vec y_i\cdot\vec r_i
\,.
	\end{equation}

The tensor matrix
element can be represented as a derivative of
the central matrix element with the same form-factor,
	\begin{equation}\label{eq:t1}
\left\langle g'\left|
f(w^\mathsf T\mathbf r)
(\mathbf y_1^\mathsf T\mathbf r)(\mathbf y_2^\mathsf T\mathbf r)
\right|g\right\rangle
=
\left(\mathbf y_1^\mathsf T\frac{\partial}{\partial\mathbf v^\mathsf T}\right)
\left(\mathbf y_2^\mathsf T\frac{\partial}{\partial\mathbf v^\mathsf T}\right)
\left\langle g'\left|f(w^\mathsf T\mathbf r)\right|g\right\rangle \,.
	\end{equation}
The central part~(\ref{eq:vtf}) of the tensor potential has a similar analytic 
representation,
	\begin{equation}\label{eq:t2}
\left\langle g'\left|
f(w^\mathsf T\mathbf r)
(w^\mathsf T\mathbf r)(w^\mathsf T\mathbf r)
\right|g\right\rangle
=
\left(w^\mathsf T\frac{\partial}{\partial\mathbf v^\mathsf T}\right)
\left(w^\mathsf T\frac{\partial}{\partial\mathbf v^\mathsf T}\right)
\left\langle g'\left|f(w^\mathsf T\mathbf r)\right|g\right\rangle \,.
	\end{equation}

Thus if the matrix element of the central potential
with a given form-factor
$f(w^\mathsf T\mathbf r)$
is
analytic, the matrix element of
the tensor potential
$f(w^\mathsf T\mathbf r)
(\mathbf y_1^\mathsf T\mathbf r)(\mathbf y_2^\mathsf T\mathbf r)$  
is also
analytic.  In particular, the tensor matrix elements of the
screened Yukawa form-factor and its descendants~(\ref{eq:sydesc}) are also
analytic.

\paragraph{Gaussian form-factor}
For a Gaussian form-factor
the matrix elements~(\ref{eq:t1},\ref{eq:t2}) are readily given as
	\begin{eqnarray}\label{eq:gt}
\left\langle g'\left|
e^{-\gamma\mathbf r^\mathsf Tww^\mathsf T\mathbf r}
(\mathbf y_1^\mathsf T\mathbf r)(\mathbf y_2^\mathsf T\mathbf r)
\right|g\right\rangle
=
\left(
\frac12\mathbf y_1^\mathsf TB'^{-1}\mathbf y_2+
(\mathbf y_1^\mathsf T\mathbf u')(\mathbf y_2^\mathsf T\mathbf u')
\right)
M'
\nonumber \,,
	\end{eqnarray}
	\begin{eqnarray}
\left\langle g'\left|
e^{-\gamma\mathbf r^\mathsf Tww^\mathsf T\mathbf r}
(\mathbf r^\mathsf T w w^\mathsf T\mathbf r)
\right|g\right\rangle
=
\left(
\frac32 w^\mathsf TB'^{-1}w + \mathbf u'^{\mathsf T} ww^\mathsf T\mathbf u'
\right)
M'
\nonumber\;,
	\end{eqnarray}
where
$B'=B+\gamma ww^\mathsf T$,
$B=A'+A$, 
$\mathbf u'=\frac12 B'^{-1}\mathbf v$,
$\mathbf v=\mathbf s'+\mathbf s$,
and where

	\begin{equation}
M'=e^{\frac14\mathbf v^\mathsf TB'^{-1}\mathbf v}\left(\frac{\pi^N}{\det(B')}\right)^{3/2}
	\end{equation}
is the central Gaussian matrix element.

Again the updates $\det(B+\gamma ww^\mathsf T)$ and $(B+\gamma ww^\mathsf
T)^{-1}$ can be efficiently calculated using rank-1 update formulas.

\paragraph{Other form-factors}

The tensor matrix element~(\ref{eq:t1}) can be written in component form as
	\begin{eqnarray}\label{eq:tg}
\left\langle g'\left|
f(w^\mathsf T\mathbf r)
(\mathbf y_1^\mathsf T\mathbf r)(\mathbf y_2^\mathsf T\mathbf r)
\right|g\right\rangle
&=&
\sum_{i,j=1}^{N}
\sum_{a,b=1}^{3}
(\vec y_1)_{ia}
(\vec y_2)_{jb}
\left\langle g'\left|
f(w^\mathsf T\mathbf r)
\vec r_{ia}
\vec r_{jb}
\right|g\right\rangle
\nonumber\\
&=&
\sum_{i,j=1}^{N}
\sum_{a,b=1}^{3}
(\vec y_1)_{ia}
(\vec y_2)_{jb}
\frac{\partial}{\partial\vec v_{ia}}
\frac{\partial}{\partial\vec v_{jb}}
\left\langle g'\left|
f(w^\mathsf T\mathbf r)
\right|g\right\rangle
\,.
	\end{eqnarray}
where $\vec r_{ia}$ is the number-$a$ component of the vector $\vec r_i$,
and where
	\begin{equation}
\left\langle g'\left|
f(w^\mathsf T\mathbf r)
\right|g\right\rangle
= MJ \,,
	\end{equation}
	\begin{equation}
M =
e^{\frac14\mathbf v^\mathsf TB^{-1}\mathbf v}
\left(\frac{\pi^N}{\det(B)}\right)^\frac32
\,,
	\end{equation}
	\begin{equation}
J =
\left(\frac\beta\pi\right)^\frac32
2\pi\frac{e^{-\beta q^2}}{\beta q}
\int_{0}^{\infty} r dr \, f(r) \, e^{-\beta r^2}\sinh(2 \beta q r) \,.
	\end{equation}

The derivative of the central matrix element at the right-hand side
of~(\ref{eq:tg}) can now be evaluated as

	\begin{eqnarray}
\frac{\partial}{\partial\vec v_{ia}}
\frac{\partial}{\partial\vec v_{jb}}
\left(
MJ
\right)
=
\left(
\frac{\partial}{\partial\vec v_{ia}}
\frac{\partial}{\partial\vec v_{jb}}
M
\right)
J
+
M
\left(
\frac{\partial}{\partial\vec v_{ia}}
\frac{\partial}{\partial\vec v_{jb}}
J
\right)
\nonumber \\
+
\left(
\frac{\partial}{\partial\vec v_{ia}}
M
\right)
\left(
\frac{\partial}{\partial\vec v_{jb}}
J
\right)
+
\left(
\frac{\partial}{\partial\vec v_{jb}}
M
\right)  
\left(
\frac{\partial}{\partial\vec v_{ia}}
J
\right)
\,,
	\end{eqnarray}
where
	\begin{equation}
\frac{\partial}{\partial\vec v_{ia}}
M
=
\sum_{k=1}^{N}
\frac12 B^{-1}_{ik}\vec v_{ka}
M
\,,
	\end{equation}
	\begin{equation}
\frac{\partial}{\partial\vec v_{jb}}
\frac{\partial}{\partial\vec v_{ia}}
M
=
\frac12 B^{-1}_{ji}\delta_{ab}
M
+
\sum_{k,l=1}^{N}
\frac12 B^{-1}_{ik}\vec v_{ka}
\frac12 B^{-1}_{jl}\vec v_{lb}
M
\,,
	\end{equation}
	\begin{equation}
\frac{\partial}{\partial\vec v_{ia}}
J
=
\frac{\partial J}{\partial q} 
\frac{\partial q}{\partial\vec v_{ia}}
\,,
	\end{equation}
	\begin{equation}
\frac{\partial}{\partial\vec v_{jb}}
\frac{\partial}{\partial\vec v_{ia}}
J
=
\frac{\partial^2 J}{\partial q^2} 
\frac{\partial q}{\partial\vec v_{jb}}
\frac{\partial q}{\partial\vec v_{ia}}
+
\frac{\partial J}{\partial q} 
\frac{\partial^2 q}{\partial\vec v_{jb}\partial\vec v_{ia}}
\,,
	\end{equation}
	\begin{equation}
\frac{\partial q}{\partial\vec v_{ia}}
=
\frac1q\sum_{k=1}^N h_i h_k \vec v_{ka}
\,,
	\end{equation}
	\begin{equation}
\frac{\partial^2 q}{\partial\vec v_{jb}\partial\vec v_{ia}}
=
\frac1q h_i h_j \delta_{ab}
-
\frac2{q^2}
\sum_{k=1}^N h_i h_k \vec v_{ka}
\sum_{l=1}^N h_j h_l \vec v_{lb}
\,.
	\end{equation}
where $h=w^\mathsf T\frac12 B^{-1}$.

Now to finish the calculation one only needs to calculate $\partial
J/\partial q$ and $\partial^2 J/\partial q^2$ for the given potential.
For the screened Yukawa form-factor and its descendants~(\ref{eq:sydesc})
the derivatives $\partial
J/\partial q$ and $\partial^2
J/\partial q^2$ are analytic although the actual calculations are relatively tedious and
should be best performed by a computer algebra software like
Maxima~\cite{maxima}. For example, the expression for
	\begin{equation}
\frac{\partial^2}{\partial q^2}
J\left[
\frac{e^{-\gamma r^2 - \mu r}}{r}
\right]
	\end{equation}
can be readily obtained by the following Maxima script,
	\begin{verbatim}
assume(beta>0,gamma>0,mu>0,q>0);
J(f) := (beta/%pi)^(3/2)*2*%pi/beta/q*exp(-beta*q^2)
*integrate(r*f*exp(-beta*r^2)*sinh(2*beta*q*r),r,0,inf);
fortran(diff(J(exp(-gamma*r^2-mu*r)/r), q, 2));
	\end{verbatim}
which analytically calculates $\frac{\partial^2 J}{\partial q^2}$ for
the screened Yukawa potential and outputs the corresponding Fortran code.

Even if $J$ is not analytic one can evaluate its derivatives
by calculating numerically a few extra integrals,

	\begin{eqnarray}
\frac{\partial J}{\partial q}=
-{{2\,\sqrt{\beta}\,e^ {- \beta\,q^2 }\,\int_{0}^{\infty }{r\,e
 ^ {- \beta\,r^2 }\,f\left(r\right)\,\sinh \left(2\,\beta\,q\,r
 \right)\;dr}}\over{\sqrt{\pi}\,q^2}}
\nonumber\\
-{{4\,\beta^{{{3}\over{2}}}\,e
 ^ {- \beta\,q^2 }\,\int_{0}^{\infty }{r\,e^ {- \beta\,r^2 }\,f\left(
 r\right)\,\sinh \left(2\,\beta\,q\,r\right)\;dr}}\over{\sqrt{\pi}}}
\nonumber\\
+
 {{4\,\beta^{{{3}\over{2}}}\,e^ {- \beta\,q^2 }\,\int_{0}^{\infty }{r
 ^2\,e^ {- \beta\,r^2 }\,f\left(r\right)\,\cosh \left(2\,\beta\,q\,r
 \right)\;dr}}\over{\sqrt{\pi}\,q}}
\,,
	\end{eqnarray}
	\begin{eqnarray}
\frac{\partial^2 J}{\partial q^2}=
{{8\,\beta^{{{5}\over{2}}}\,e^ {- \beta\,q^2 }\,\int_{0}^{\infty }{
 r^3\,e^ {- \beta\,r^2 }\,f\left(r\right)\,\sinh \left(2\,\beta\,q\,r
 \right)\;dr}}\over{\sqrt{\pi}\,q}}
\nonumber\\
+{{8\,\beta^{{{5}\over{2}}}\,q\,e
 ^ {- \beta\,q^2 }\,\int_{0}^{\infty }{r\,e^ {- \beta\,r^2 }\,f\left(
 r\right)\,\sinh \left(2\,\beta\,q\,r\right)\;dr}}\over{\sqrt{\pi}}}
\nonumber\\
+
 {{4\,\beta^{{{3}\over{2}}}\,e^ {- \beta\,q^2 }\,\int_{0}^{\infty }{r
 \,e^ {- \beta\,r^2 }\,f\left(r\right)\,\sinh \left(2\,\beta\,q\,r
 \right)\;dr}}\over{\sqrt{\pi}\,q}}
\nonumber\\
+{{4\,\sqrt{\beta}\,e^ {- \beta\,q
 ^2 }\,\int_{0}^{\infty }{r\,e^ {- \beta\,r^2 }\,f\left(r\right)\,
 \sinh \left(2\,\beta\,q\,r\right)\;dr}}\over{\sqrt{\pi}\,q^3}}
\nonumber\\
-{{8\,
 \beta^{{{3}\over{2}}}\,e^ {- \beta\,q^2 }\,\int_{0}^{\infty }{r^2\,e
 ^ {- \beta\,r^2 }\,f\left(r\right)\,\cosh \left(2\,\beta\,q\,r
 \right)\;dr}}\over{\sqrt{\pi}\,q^2}}
\nonumber\\
-{{16\,\beta^{{{5}\over{2}}}\,e
 ^ {- \beta\,q^2 }\,\int_{0}^{\infty }{r^2\,e^ {- \beta\,r^2 }\,f
 \left(r\right)\,\cosh \left(2\,\beta\,q\,r\right)\;dr}}\over{\sqrt{
 \pi}}}
\,.
	\end{eqnarray}

\subsubsection{Spin-orbit potential}

The spin-orbit potential between two particles with coordinates
$\vec r_1$ and $\vec r_2$ and spins $\vec S_1$ and $\vec S_2$ can be
written in the form~\cite{ring}

	\begin{equation}
V_\mathsf{so} \propto f(r)\left(\vec S\cdot\vec L\right) \,,
	\end{equation}
where
$\vec r=\vec r_1-\vec r_2$ is the relative coordinate between the particles;
$f(r)$ is the radial form-factor of the potential;
$\vec S=\vec S_1+\vec S_2$
is the total spin of the two particles;
and $\vec L$ is the relative orbital momentum between the two particles,
	\begin{equation}
\vec L=
(\vec r_1-\vec r_2)\times
\frac{-i}{2}\left(\frac{\partial}{\partial\vec r_1}-\frac{\partial}{\partial\vec r_2}\right) \,,
	\end{equation}
where ``$\times$'' denotes vector-product of two vectors.

The orbital momentum operator can be written, using the size-$N$ column of numbers $w$,
	\begin{equation}
w = \{w_i | w_1=1, w_2=-1, w_{i\ne 1,2}=0\} \,,
	\end{equation}
in the general form,
	\begin{equation}
\vec L = \frac{-i}{2}\left(
w^\mathsf T\mathbf r\times w^\mathsf T\frac{\partial}{\partial\mathbf r^\mathsf T}
\right) \,,
	\end{equation}
where
	\begin{equation}
w^\mathsf T\mathbf r\equiv \sum_{i=1}^{N}w_i\vec r_i \,,\;
w^\mathsf T\frac{\partial}{\partial\mathbf r^\mathsf T}\equiv
\sum_{i=1}^{N}w_i\frac{\partial}{\partial\vec r_i}
\,.
	\end{equation}

For a given form-factor $f(w^\mathsf T\mathbf r)$ the spin-orbit matrix
element can be represented through the central matrix element with the
same form-factor,

        \begin{eqnarray}\label{eq:mso}
\left\langle g'\left|
f(w^\mathsf T\mathbf r)
\left(
w^\mathsf T\mathbf r\times
w^\mathsf T\frac{\partial}{\partial\mathbf r^\mathsf T}
\right)
\right|g\right\rangle
=
\left(w^\mathsf T\frac{\partial}{\partial\mathbf v^\mathsf T}\right)
\times w^\mathsf T
\left(
\mathbf s - 2A\frac{\partial}{\partial\mathbf v^\mathsf T}
\right)
\left\langle g'\left|
f(w^\mathsf T\mathbf r)
\right|g\right\rangle
\,,
        \end{eqnarray}
so that if the central matrix element is analytic --- as is the case for
the screened Yukawa form-factor and its descendants~(\ref{eq:sydesc})
--- the spin-orbit matrix element is also analytic.

\paragraph{Gaussian form-factor}
For a Gaussian form-factor
the spin-orbit matrix element is given as
	\begin{eqnarray}
\left\langle g'\left|
e^{-\gamma\mathbf r^\mathsf Tww^\mathsf T\mathbf r}
\left(
w^\mathsf T\mathbf r\times
w^\mathsf T\frac{\partial}{\partial\mathbf r^\mathsf T}
\right)
\right|g\right\rangle
=
\frac12 w^\mathsf TB'^{-1}\mathbf v\times
\left(w^\mathsf T\mathbf s-w^\mathsf TAB'^{-1}\mathbf v\right)
M'
\,,
	\end{eqnarray}
where
$B'=B+\gamma ww^\mathsf T$,
$B=A'+A$,
$\mathbf u'=\frac12 B'^{-1}\mathbf v$,
$\mathbf v=\mathbf s'+\mathbf s$,
and where

        \begin{equation}
M'=e^{\frac14\mathbf v^\mathsf TB'^{-1}\mathbf v}\left(\frac{\pi^N}{\det(B')}\right)^{3/2}
        \end{equation}
is the central Gaussian matrix element.

Again $\det(B')$ and ${B'}^{-1}$ can be efficiently calculated using rank-1 update formulas.

\paragraph{General form-factor}
The component number~$a$ of the spin-orbit matrix element~(\ref{eq:mso})
can be written in component form as

	\begin{eqnarray}
\left\langle g'\left|
f(w^\mathsf T\mathbf r)
\left(
w^\mathsf T\mathbf r\times
w^\mathsf T\frac{\partial}{\partial\mathbf r^\mathsf T}
\right)_a
\right|g\right\rangle
=
\\
\sum_{b,c=1}^3 \epsilon_{abc}
\sum_{k=1}^N w_k \frac{\partial}{\partial\vec v_{kb}}
\sum_{l=1}^N w_l \left(
\vec s_{lc}-2\sum_{j=1}^N A_{lj}\frac{\partial}{\partial\vec v_{jc}}
\right)
\left\langle g'\left|
f(w^\mathsf T\mathbf r)
\right|g\right\rangle
\,.
	\end{eqnarray}
It is clearly a linear combination of the first,
$\frac{\partial}{\partial\vec v_{kb}}
\left\langle g'\left| f(w^\mathsf T\mathbf r) \right|g\right\rangle$,
and second,
$\frac{\partial}{\partial\vec v_{kb}}\frac{\partial}{\partial\vec v_{jc}}
\left\langle g'\left| f(w^\mathsf T\mathbf r) \right|g\right\rangle$,
derivatives of the corresponding central matrix element
$\left\langle g'\left| f(w^\mathsf T\mathbf r) \right|g\right\rangle=MJ$.
These quantities have been calculated in the previous chapter.

\subsection{Many-body forces}
Many-body potentials in nuclear physics have form-factors which
depend on the coordinates of several nucleons.  In addition they
may have tensor and spin-orbit post-factors.

For a many-body potential with a Gaussian form-factor, $e^{-\mathbf r^\mathsf
T W \mathbf r}$,  where $W$ is a symmetric positive-definite matrix,
the central matrix element is given as
	\begin{equation}
\left\langle g'\left|
e^{-\mathbf r^\mathsf T W \mathbf r}
\right|g\right\rangle
=
e^{\frac14\mathbf v (B+W)^{-1} \mathbf v}
\left(
\frac{\pi^N}{\det(B+W)}
\right)^{3/2}
\,.
	\end{equation}

The simple tensor and spin-orbit post-factors can be calculated in the
same way as has been done in the previous two chapters.

\section{Conclusion}

In quantum few-body physics the analytic matrix elements are of importance
for the correlated Gaussians method as they facilitate extensive numerical
optimizations of the variational wave-functions.  In this paper it has
been shown that potentials in the form

	\begin{equation}
f(w^\mathsf T\mathbf r) \,,\;
f(w^\mathsf T\mathbf r)(\mathbf y^\mathsf T\mathbf r)(\mathbf z^\mathsf T\mathbf r) \,,\;
f(w^\mathsf T\mathbf r)(a^\mathsf T\mathbf r)(b^\mathsf T\mathbf r) \,,\;
f(w^\mathsf T\mathbf r)
\left(
w^\mathsf T\mathbf r \times w^\mathsf T\frac{\partial}{\partial\mathbf r^\mathsf T}
\right) \,,
	\end{equation}
have analytic matrix elements between shifted correlated Gaussians
for the following class of form-factors,
	\begin{equation}
f(r)=r^n \frac{e^{-\gamma r^2-\mu r}}{r} \,,\;n=0,1,\dots
\,.
	\end{equation}
Of these form-factors the Gaussian form-factors produce
particularly simple and concise analytic expressions.
Therefore an efficient strategy could be to represent the potentials
at hand as linear combinations of Gaussians and then use the analytic
expressions for the Gaussians.


\begin{thebibliography}{10}

\bibitem{mitroy} J.~Mitroy et al., Theory and applications of explicitly
correlated Gaussians, Rev. Mod. Phys. 85, 693 (2013).

\bibitem{ChemRev} Sergiy Bubin, Michele Pavanello, Wei-Cheng Tung, Keeper
L. Sharkey, and Ludwik Adamowicz, Born-Oppenheimer and Non-Born-Oppenheimer Atomic and
Molecular Calculations with Explicitly Correlated Gaussians, Chemical Reviews 113, 36 (2013).

\bibitem{cafiero-analytic} Mauricio Cafiero, Ludwik Adamowicz, Analytical Gradients for Singer’s
Multicenter n-Electron Explicitly
Correlated Gaussians, International Journal of Quantum Chemistry, Vol. 82, 151–159 (2001).

\bibitem{sharkey-analytic} Keeper L. Sharkey, Sergiy Bubin, and Ludwik Adamowicz,
An algorithm for calculating atomic D states with explicitly correlated
Gaussian functions, The Journal of Chemical Physics 134, 044120 (2011).

\bibitem{DailyGreene} K. M. Daily and Chris H. Greene, Extension of
the correlated Gaussian hyperspherical method to more particles and
dimensions, Physical~Review~A~89, 012503 (2014)

\bibitem{tung-calc} Wei-Cheng Tung and Ludwik Adamowicz, Accurate potential
energy curve of the LiH+ molecule calculated with explicitly correlated
Gaussian functions, J. Chem. Phys. 140, 124315 (2014).

\bibitem{bubin-calc} Sergiy Bubin, Martin Formanek, Ludwik Adamowicz, Universal
all-particle explicitly-correlated Gaussians for non-Born–Oppenheimer
calculations of molecular rotationless states, Chemical Physics Letters 647, 122 (2016).

\bibitem{blume-semi} X.Y. Yin and D. Blume, Trapped unitary two-component
Fermi gases with up to ten particles, Physical Review A 92, 013608 (2015).

\bibitem{gsl} M. Galassi et al, GNU Scientific Library Reference Manual -
Third Edition (January 2009), ISBN 0954612078.

\bibitem{suzukivarga} Y.~Suzuki and K.~Varga, Stochastic Variational Approach to
Quantum-Mechanical Few-Body Problems, ISBN 3-540-65152-7, Springer-Verlag,
Berlin, 1998.

\bibitem{ring} Peter~Ring and Peter~Schuck, The Nuclear Many-Body Problem,
ISBN 3-540-09820-8, Springer-Verlag, Berlin, 1980.

\bibitem{maxima} Maxima.sourceforge.net. Maxima, a Computer Algebra
System. Version 5.34.1 (2014). http://maxima.sourceforge.net/

	\end{thebibliography}
\end{document}